
\def\log{\rm log}
\def\rs{{$\phantom {[123]}$}}
\def\today{\ifcase\month
\or January\or February\or March\or April\or May
\or June\or July\or
August\or September\or October\or November\or December\fi
\space\number\day, \number\year}
\tenrm 
\font\sect=cmr10 scaled \magstep2  
\font\text=cmr10  
\baselineskip14pt
\voffset=2\baselineskip
\parskip=.1truecm
\raggedbottom
\footline={\ifnum\pageno=0{}\else\hss\tenrm\folio\hss\fi}
\pageno=0
\hsize=15truecm
\vsize=22truecm
\def\secttitle #1
{\hbox{ }\vskip.5truecm
\par{\bigbreak\noindent{\sect #1}\par\medbreak}\nobreak\text}
\def\next{\hfil\break}
\def\rs{$\phantom {12345}$}
\hbox {}
\font\tittlefont=cmr10 scaled \magstep2
\font\names=cmr10
\vskip 2truecm
\centerline {\tittlefont Parallel P3M with exact calculation of short range
forces}
\vskip 1truecm
\noindent\names
Subject headings: Astrophysics\next
\vskip 3truecm

\names
\centerline {T. Theuns$^1$}
\centerline {Scuola Normale Superiore}
\centerline{Piazza dei Cavalieri 7,
I-56126 Pisa,
Italy}
\centerline {THEUNS\%vaxsns.dnet@ux1sns.sns.it}
\smallskip

\vskip 4truecm
\noindent
Accepted for publication by Computer Physics Communications, September
24, 1993\next
\smallskip
\noindent $^1$ Present Address:\next
\item {} Department of Physics
\item {} Astrophysics
\item {} Nuclear Physics Laboratory
\item {} Keble Road Oxford
\item {} OX1 3RH
\item {} UK
\item {} tt@oxds02.astro.ox.ac.uk

\vfill\break

\secttitle {Abstract}

A P3M (Particle-Particle, Particle-Mesh) algorithm to compute the
gravitational force on a set of particles is described. The
gravitational force is computed using Fast Fourier Transforms. This
leads to an incorrect force when the distance between two particles is
of the order of a grid cell. This incorrect force is subtracted
exactly from all particles in parallel using convolution with the
appropriate Green's function in real space in a time of order $N_T$,
irrespective of the degree of clustering of particles.  Next, the
correct $1/r^2$ force is added for all neighbouring particles in
parallel, leading to an accurate algorithm which runs efficiently on a
highly parallel computer. A full force calculation for $128k$
particles on a $128^3$ grid in a mildly clustered situation requires
approximately 196 seconds on a $8k$ Connection Machine 2 with 8MHz
clock.  This decreases to an estimated 9.8 seconds on a full-sized
$64k$ CM200.

\vfill\break

\secttitle {1. Introduction}

In many fields of computational astronomy it is necessary to compute
the gravitational force on a set of massive particles. The optimum
algorithm to do this depends on what one wants to learn from the
results.

The particle-particle (PP, or direct summation) method (Aarseth [1])
gives the highest accuracy, necessary for simulations where encounters
between particles are real and are to be calculated accurately. (Such
is the case in for example star cluster simulations.) The method is
expensive since computer time $\tau$ scales rather steeply with
particle number $N_T$ : $\tau \propto N_T^{1.6}$, where the exponent
depends somewhat on the actual particle distribution computed upon.
The PP scheme can be implemented very efficiently on many computer
architectures.

The particle-mesh (PM) method (e.g. Hockney and Eastwood [2], chapter
5) uses a grid to compute interactions between particles in three
steps: (1) compute a density distribution on the grid that faithfully
represents the density distribution of the particles, (2) solve
Poisson's equation on the grid and compute the associated grid forces
and (3) interpolate these forces back to the particles. Operations
involving particles occur only in steps (1) and (3) and they scale
linearly with $N_T$. The solution of Poisson's equation in step (2) is
independent of $N_T$. All three steps require many fewer operations
than the PP method and consequently many more particles can be used in
a simulation for the same amount of CPU time spent. Unfortunately,
this grid-mediated force $f$ computed between two particles which are
only a few grid cells away from each another is inaccurate: it is both
anisotropic and not translationally invariant.  Consequently, the
method is useful when the force on a particle is mainly determined by
the smooth density distribution of all other (distant) particles--
that part of the force being calculated accurately, and not so much by
the irregular distribution of near neighbours (for example the force
on a star in a galaxy).  The solution of Poisson's equation in step 2
can be done very efficiently using Fast Fourier Transforms (FFT's), an
algorithm which many computer vendors supply and which runs at near
maximum machine speed.  The PM method allows for a trivial
implementation of periodic boundary conditions.  Isolated boundary
conditions can also be treated using FFT's, but at rather a high cost
(James [3]).

The particle-particle, particle-mesh (P3M, Hockney and Eastwood [2],
chapter 8, Efstathiou et al. [4]) tries to combine the best of both PP
and PM methods, namely use FFT's on a grid to compute forces between
distant particles efficiently (PM step) and use direct summation to compute
forces between near particles accurately (PP step).  Hence, forces are
computed in three steps: (1) the PM step, to compute forces between
distant particles using FFT's, (2) a correction step, to subtract the
inaccurate PM force for particles close to one another and (3) the PP
step, to add the correct $1/r^2$ force between those near neighbours.
The correction force subtracted in step 2 is computed as a function of
interparticle distance using the Green's function used to solve
Poisson's equation on the grid.  However, the applied correction is
not exact since this grid-force is neither isotropic nor
translationally invariant which means that the correction is not a
function of interparticle distance alone. The Green's function is
smoothed on small scales to decrease the difference between correction
term and grid-force in the least squares sense. This algorithm has the
advantage over the PM one in that the resolution of the calculation is
not limited by the size of the grid cells, an important factor in
simulations where the amount of clustering is not constant in time
(e.g. in simulations of large scale structure formation). Both the
correction force in step 2 and the PP force in step 3 are calculated
on a per pair basis, hence required CPU time scales with $N_b^2$ for
both, where $N_b$ is the average number of neighbours to be
corrected. This leads to a slowdown of the speed of the
program as the amount of clustering increases. This undesirable
feature may be partly compensated for by introducing multiple grids
(Villumsen [5]).

The hierarchical tree algorithm (Appel [6], Barnes and Hut [7]) groups
distant particles and uses a multipole expansion to decrease the
number of terms occurring in the calculation of the force on one
particle from $N_T$ to log$N_T$.  Forces between neighbouring particles
are computed on a particle-particle basis. The algorithm allows the
use of individual time-steps which improves its efficiency
substantially in some situations. The tree traversal algorithm can be
vectorized efficiently on architectures favouring small vectors
(Hernquist [8]), however, such codes run rather inefficient on highly
parallel computers that require long vectors (Makino and Hut [9]).

In this paper we discuss a P3M implementation that differs on two
grounds from the P3M discussed in Efstathiou et al. [4]: (1) the
method uses convolution in real space to subtract the grid-mediated
force on neighbour particles exactly in a time of order $N_T$,
irrespective of the particle distribution and (2) sorts particles in
cells to efficiently compute the PP part on a highly parallel computer
(in our case on a 8192 nodes Connection Machine 2 (CM2)). Both the PM
part, using FFT's and the PP part, using a scan-copy approach, run
very efficiently on the CM2. The convolution part that corrects the
short range force is described in section 2.  The parallel PP part is
described in section 3. Final remarks can be found in section 4.
The Appendix describes some properties of the CM2 and in addition
gives some optimization and algorithmic design considerations for the
operations needed in both the PM and PP calculations.

\secttitle{2. Correcting the short range force}

The computation of the grid-mediated interaction consists of the
following steps (see e.g. Hockney and Eastwood [2], chapter 5 or
Efstathiou et al. [4] for more details): (1) given the distribution of
particles, compute the density on the grid, (2) compute the grid
potential by convolving the density distribution with the Green's
function and difference it to obtain the grid forces, and (3)
interpolate these forces to the particles. In the following we assume
that the assignment of mass to the grid and of forces to particles
uses cloud-in-cell assignment (i.e.  linear interpolation in all three
Cartesian directions) but other assignment schemes can be considered
as well.

The density $\rho(i,j,k)$ at vertex $(i,j,k)$ of the 3D grid due to
particles $n=1\dots N_T$ with masses $m(n)$ and coordinates
$(x(n),y(n),z(n))$ is:
$$\rho(i,j,k) = {1\over V_c} \sum_{n=1}^{N_T} m(n)
f(X_i,x(n)) f(Y_j,y(n)) f(Z_k,z(n)),\eqno(1)$$ where we have defined:
$$f(x,y) = \cases {0;&if $|x-y|>c$;\cr
1-|x-y|/c,&otherwise.\cr}\eqno(2)$$ Here, $c$ is the size of a grid
cell, $V_c=c^3$ its volume and $(X_i, Y_j, Z_k)$ its Cartesian
coordinates.  Given this density distribution, one next calculates the
potential $\Phi(i,j,k)$ by convolving $\rho$ with the Green's function
$G$:
$$\Phi(i,j,k) = \sum_{p,q,r} \rho (p,q,r) G(i-p,j-q,k-r).\eqno(3)$$
The convolution in Eq. (3) can be done very efficiently by means of
FFT's. Given the potential $\Phi$ one needs to compute the forces on
the grid. This can be done, for example, using the following centered
difference
expression:
$$FX(i,j,k) = (\Phi(i+1,j,k) -\Phi(i-1,j,k)) / 2c.\eqno(4)$$
Finally, these forces are interpolated
back to the particles, using an equation similar to Eq. (1):
$$Fx(n) = \sum_{i,j,k} FX(i,j,k) f(X_i,x(n)) f(Y_j,y(n) f(Z_k,z(n)).
\eqno(5)$$
Note that a particle $n$ receives forces from the eight vertices nearest
to it in Eq. (5), just as it contributes mass to only eight vertices in
Eq. (1). This completes the calculation of the PM forces.

The force $Fx(n_1)$ on a particle $n_1$ will have a contribution to it
from a particle $n_2$, say. If the distance between $n_1$ and $n_2$ is
small compared to the cell size $c$, then this contribution is a poor
approximation to the correct force. However, we now show that we can
use equations (1)-(5) to compute these inaccurate forces due to
particles close to one another.  Once the inaccurate interparticle
force is subtracted, one can add the correct force on $n_1$ due to
$n_2$. (The latter PP calculation is described in section 3.) It is
convenient to subtract forces from particles $n_2$ which are either in
the same cell as $n_1$ or in one of the (26, in 3D) neighbouring cells.
The algorithm works for particles any number of cells away from $n_1$,
but the amount of computer time needed for the subtraction of the
inaccurate force and especially for the addition of the correct force
in the PP part becomes rapidly prohibitively large.

Assume for the purpose of explanation that the system is one
dimensional and number the vertices of the grid $1\cdots N_g$.  We
will say that a particle is in cell $i$ if the left vertex of the cell
this particle is in, is $i$. Particles in cell $i$ receive forces from
vertices $i$ and $i+1$.  In turn, the force $FX(i)$, on vertex $i$,
depends on the potential on vertices $i-1$ and $i+1$. Since we want to
correct the force on vertex $i$ and $i+1$ due to particles in cells
$i-1$, $i$ and $i+1$, we need to calculate the potentials on vertices
$i-1,\cdots, i+2$, due to those particles.

For this purpose, calculate the quantity $\rho_L(i)$ defined
by
$$\rho_L(i) = {1\over V_c}\ \sum_{n=1}^{N_T} m(n) g(X_i,x(n)),\eqno(6)$$
where
$$g(x,y) = \cases {0,&if $|x-y|> c$, or $x>y$;\cr 1-|x-y|/c
&otherwise.\cr}\eqno(7)$$ Comparing Eq. (6) with Eq. (1), it is clear
that $\rho_L(i)$ is the density at vertex $i$ due to all particles in
the cell $i$ (the left ($L$) vertex), whereas $\rho$ (defined by Eq.
(1) for the 3D case) is the density at
vertex $i$ due to particles in cells $i-1$ and $i$. In an analogue way
we compute the density $\rho_R(i)$, which is the density at vertex
$i+1$ due to particles in cell $i$.  (Consequently, $\rho(i) =
\rho_L(i)+\rho_R(i-1)$.) Given $\rho_L$ and $\rho_R$, we
compute the densities $\rho_{-1}$, $\rho_0$, $\rho_1$ and $\rho_2$
defined by:
$$\eqalignno{
\rho_{-1}(i) & = \rho_L(i-1) &(8.1)\cr
\rho_0(i)    & = \rho_R(i-1) + \rho_L(i)(\equiv \rho(i))&(8.2)\cr
\rho_1(i)    & = \rho_R(i) + \rho_L(i+1)(\equiv \rho(i+1))&(8.3)\cr
\rho_2(i)    & = \rho_R(i+1).&(8.4)\cr
}$$ These densities are the ones that determine the forces in cell $i$
due to particles in cells $i-1$, $i$ and $i+1$.  We then use these
densities to compute the potentials $\Phi_{-1}(i),\cdots ,\Phi_2(i)$
using the convolution in real space:
$$\Phi_l(i) = \sum_{m=-1}^2
\rho_{m}(i) G(l-m),\eqno(9)$$ for $l=-1,\cdots ,2$. Note that the
convolution (9) contains only four terms whereas the full convolution
(3) contains $N_g$ terms (in 1D). $G$ denotes the Green's function in
real space, as before.  Finally, we compute the forces $FX_L(i)$ and
$FX_R(i)$ as
$$\eqalignno{ FX_L(i) & = (\Phi_1(i)-\Phi_{-1}(i)) / 2c&(10.1)\cr
FX_R(i) & = (\Phi_2(i)-\Phi_{0}(i))/ 2c, &(10.2)\cr }$$ which are the
correction forces to be interpolated to the particles in cell $i$ from
the left ($L$) and right ($R$) vertices of cell $i$ respectively.

The forces between two particles, obtained using this procedure, are
exactly equal to the forces computed using the PM algorithm if these
particles are either in the same or in neighbouring cells. If
particles are further away from each other, then these forces are
exactly zero.

Our implementation uses circular shifting (see Appendix) of the grid
arrays (like e.g. $\rho_L$ and $\rho_R$) in all three Cartesian
directions to compute the neighbour forces to subtract.  This has the
advantage that the same boundary conditions are imposed in PM and
correction part if the same grid is used in both parts (i.e. if the
FFT computes the forces for periodic boundary conditions, then also
the subtraction of neighbour forces will be done as if the system were
periodic). The same is true for isolated boundary conditions.

The computational expense of this correction step is not dramatic. In
a 3D system, one has to compute corrections from all eight vertices of
the cell the given particle is in, hence the force assignment requires
the same amount of computation as in the PM part: $\approx
9*3*8N_T=216N_T$ operations. However, for each of these eight vertices
one needs to compute corrections due to all eight vertices a particle
assigns mass to, i.e., the mass assignment needs to be done eight
times more often than in the PM part. (This can be reduced, but then
one needs to introduce many more temporary variables to store the grid
density for all eight vertices separately, in the 1D example one needs
to store $\rho_L$ and $\rho_R$ separately, whereas in the PM part one
needs to store only the sum $\rho_L(i)+\rho_R(i-1)$.) Mass assignment
takes $\approx 11*8N_T=88N_T$ operations.  One convolution like Eq.~
(9), requires many fewer operations than the PM part: as Eq.~(9)
shows, one needs to multiply the densities $\rho$ with the central
four values of the Green's function.  These parallel statements
require computer time scaling linearly with the total number of
vertices and in addition are executed at near top speed on a parallel
computer since the amount of communication required to complete them
is virtually zero. (The PM convolution Eq. (3) requires of order
$21N_g^3\log N_g^3$ operations.) On the other hand, one needs to
perform the convolution (9) 1728 times in order to correct for all
contributing vertices, thus making this part of the algorithm
responsible for most of the CPU time needed for the correction step,
notwithstanding the extremely high execution speed with which Eq.~(9)
is computed (close to 500 MFlops on the 8k CM2). We will show in
section 3 that the PP calculation requires a comparable amount of CPU
time.

Some timings of the PM and correction algorithm on a 8192 processor
CM2 running at 8 MHz are given in Table 1. These timings were obtained
using the CM2 vendor supplied timer and are reproducible to within
10\%.  The CPU time required is given for the mass and force
assignments during the PM step (column 3), the FFT step (column 4),
the mass assignment and force assignment part in the correction step
(column 5), the convolution of Eq. (9) (column 6) and the total
calculation (column 7) for the given number of particles $N_T$ ($k$
stands for 1024) and grid vertices $N_g$ (in 1D, i.e. the timed 3D
system has $N_g^3$ vertices). The time stated for the FFT part
includes the computation of the Green's function and the extraction of
its central region, needed in the convolution part of the correction
step, in addition to the actual FFT calculation. We reiterate that all
these times are independent of the actual particle distribution.

\vbox{\smallskip\centerline{\bf Table 1}
\vskip10pt
\centerline{\bf CPU time (seconds) for PM calculation and correction
step}
\tabskip=1.5em plus1em minus1em
\halign to\hsize{\hfil #\hfil &\hfil #\hfil&\hfil #\hfil & \hfil #\hfil &\hfil
#\hfil\ &\hfil #\hfil&\hfil #\hfil\cr
\noalign{\vskip10pt\hrule\vskip6pt\hrule\vskip6pt}
$N_T$&$N_g$&PM&FFT&Correction &Convolution&Total\cr
&&assignments&&assignments&\cr
8k	&32	&0.2	&0.12	&0.8	&1.8	&3.0\cr
64k	&32	&0.9	&0.12	&5.5	&1.8	&7.3\cr
64k	&64	&1.4	&0.93	&6.0	&10.8	&19.1\cr
64k	&128	&6.2	&7.26	&16.7	&73.2	&103.4\cr
128k	&128	&6.9	&7.26	&22.1	&73.2	&109.5\cr
\noalign{\vskip6pt\hrule\vskip6pt}}\smallskip}

Figure 1 illustrates the accuracy of the PM force obtained in the way
just described. The figure shows the deviation of the radial force,
the tangential force and the potential from their Newtonian values for
randomly distributed massless test particles in a $16^3$ grid with
unit cell size. The forces are due to eight equal mass particles that
were also randomly distributed in cell (8,8,8).  The truncated Green's
function appropriate for an isolated system was used and the PM FFT
calculation was computed on a grid of doubled size (in each direction)
to avoid aliasing, as described in Hockney and Eastwood ([2], p.~
213). Deviations of the force of order 6\% occur up to six cells away,
and the force never quite reaches it's $1/r^2$ behaviour (a deviation
$\le 1$\% remains).  The latter is probably due to differencing errors
in obtaining the force from the potential since the potential is more
accurate than the force, as Fig. 1 shows.

In summary, the result of the two steps, the PM and the correction
step is the following: for two particles $n_1$ and $n_2$ more than two
cells apart, the force is computed using the PM algorithm. If these
particles are in the same cell or in neighbouring cells, the force
between them is exactly zero.  In the next section, we describe a
parallel algorithm to compute the forces due to these neighbouring
particles.

\secttitle{3. Parallel Particle-Particle part}

The PP part becomes increasingly time-consuming as the degree of
clustering of particles increases and consequently more PP forces need
to be computed. Hence it is important that the algorithm deals
efficiently with such clustered particle distributions.  The algorithm
we propose is a variant of that discussed in Theuns and Rathsack
([10]), in the context of a parallel SPH (Smoothed Particle
Hydrodynamics) implementation. It uses large vectors and a minimum of
inter-processor communication to run efficiently on a massively
parallel computer like the CM2. In addition, it is
written in such a way that densely populated cells interact very
efficiently with sparsely populated ones.

Given the basic requirements of the CM2 as discussed in the Appendix,
the main idea of the algorithm is to construct two vectors, a Bottom
($B$) vector and a Top ($T$) vector, which are organized such that
interaction can occur between a particle $B(i)$ and a particle $T(i)$.
Such an interaction can be computed very efficiently on a parallel
computer since $B(i)$ and $T(i)$ are in the private memory of {\it the
same} processor. After this calculation, $T$ is reorganized so that
$B(i)$ interacts with another particle in $T$ and these two steps are
repeated until all interactions have been computed. The task is now to
make this reorganization as efficient as possible.

The algorithm we use is different for interactions between particles
in the same cell and interactions between particles in neighbouring
cells. In the first step, we want to calculate all interactions
between a particle and all other particles {\it in the same} cell. We
do this by loading all particles in $B$ such that particles in the
same cell are in consecutive array locations (which we will call \lq
segments\rq~). $T$ starts out as being a copy of $B$ which we CSHIFT
by one to the left (see the Appendix which discusses CSHIFT): every
particle $B(i)$ has now a (different) particle $T(i)$ above it which
is in the same cell and so we can compute their mutual interaction. We
then CSHIFT $T$ left one more step and compute another interaction. We
repeat the process until all interactions have been computed. Remark
that, as the process advances, particles will be shifted into $T(i)$
which do {\it not} belong to the same cell as $B(i)$, since not all
cells have the same number of particles in them.  Consequently, the
algorithm becomes less and less efficient in calculating forces as
more and more cells have been processed and valid interactions occur
in only the most populated cells. Figure 2a illustrates this processing
of intra-cell interactions.

Next we describe how to compute interactions between cells.  Again we
load particles into segments in array $B$.  The segments themselves
are organized in such a way that particles of two cells that interact
are in the same locations in $B$ and $T$, i.e., if a cell is found in
locations $B[i_1:i_n]$ then the particles of a neighbour cell are in
locations $T[i_1:i_m]$. Of the interacting cells, we always send the
larger one to $B$, i.e. $n\ge m$. Array $B$ will contain of order 13/2
times the number of particles, since each cell interacts with 13
neighbouring cells (because of symmetry we don't need to process all
26 neighbouring cells).  The algorithm now proceeds as follows: (1)
copy the first particle of each of the segments in $T$ to all
particles in its corresponding segment in $B$, (2) compute the
interactions between this particle and all particles in the bottom
segment, (3) sum the force contributions of the copied particle along
the segment and store the sum, (4) shift the particles in $T$ by one
position to align a new, unprocessed particle of $T$ with its
corresponding segment in $B$, (5) repeat steps (1)-(4) until all
interactions have been computed. Finally, the results of all 13
cellular interactions are added to the force due to
particles in the same cell to obtain the neighbour force for each
particle. The copying in step (1) and the summing in step (3) are done
using special routines called scanning routines, supplied by the
vendor (see Appendix). Figure 2b illustrates this part of the
algorithm.

We want to stress the high level of parallelization thus obtained: all
cells are processed in parallel in the intra-cell calculation and in
addition, all 13 directions are processed in parallel during the
inter-cell calculation. The copying of particles from $T$ along
segments in $B$ has the advantage that the time required to compute
the forces depends only on the maximum number of particles in a
segment in $T$ and not on the number of particles in segments of $B$.
Since we take care always to put the densest cells in $B$ we can also
compute {\it uneven} particle distributions rather efficiently.

Table 2 presents timings of the PP algorithm to illustrate the overall
efficiency. The triply periodic system which was timed was set-up in
the following way: $N_T$ particles were distributed in cells using a
random number generator such as to give a prescribed value of the
dispersion of cell occupation numbers
$\sigma\equiv<(N_c-<N_c>)^2>^{1/2}/<N_c>$, where $<N_c>$ is the
average number of particles per cell. Table 2 gives the values for
$N_T$, $\sigma$ and the maximum number of particles in a given cell,
$N_{max}$ in columns 1-3. The next four columns give the required
CPU-time (in seconds) for (4) the organization associated with the
intra-cell computation, (5) the intra-cell computation itself, (6) the
organization associated with inter-cell communication and (7) the
inter-cell computation itself.  The $32k$ system consisted of $16^3$
cells and so had $<N_c>$ = 8, while the $128k$ system had $32^3$ cells
and $<N_c>$ = 4. We want to stress that in this part of the
calculation, computer time is {\it less} when more vertices $N_g$ are
used, since fewer PP forces need to be computed in this case.

The intra-cell calculation scales linearly with $N_T\ N_{max}$, since
the required time is proportional to both the maximum number of
particles in a cell and the total number of particles. The time
required to compute the inter-cell interactions increases with
increasing clustering $\sigma$ at first but then
becomes nearly constant, because fewer cells take part in the
calculation with increasing clustering at that stage.  This is
probably pathological to our chosen particle distribution, so we
generated another particle distribution where each cell contained at
least one particle.

\vbox{
\centerline{\bf Table 2}
\vskip10pt
\centerline{\bf CPU-usage (seconds) for PP part}
\tabskip=1.5em plus1em minus1em
\halign to\hsize{\hfil #\hfil &\hfil #\hfil&\hfil #\hfil & \hfil #\hfil &\hfil
#\hfil\ &\hfil #\hfil&\hfil #\hfil\cr
\noalign{\vskip10pt\hrule\vskip6pt\hrule\vskip6pt}
$N_T$&$\sigma$&$N_{max}$&set-up
1&intra-cell&set-up 2&inter-cell\cr
\noalign{\vskip6pt\hrule\vskip6pt}
32k&0.0&	8&	0.22&	0.04&		2.5&		3.6\cr
32k&0.9&	21&	0.22&	0.13&		2.3&		7.2\cr
32k&2.1&	65&	0.25&	0.38&		1.8&		10.8\cr
32k&4.9&	294&	0.27&	1.75&		1.0&		10.5\cr
\cr
128k&0.0&	4&	1.40&	0.07&		12.2&		10.7\cr
128k&0.9&	10&	0.8&	0.19&		11.3&		15.8\cr
128k&2.1&	33&	0.8&	0.54&		8.4&		13.1\cr
128k&4.1&	106&	0.8&	2.3&		7.1&		15.9\cr
\noalign{\vskip6pt\hrule}
}
\smallskip}

Results for this distribution are presented in Table 3 and illustrate
the load-balancing problem that occurs for this more realistic
particle distribution. The first set of four lines are appropriate for
the algorithm as described so far, the next four are for the
load-balanced algorithm to be described next, using the same particle
distribution.  Table 3 shows that the old algorithm becomes very much
less efficient as the degree of clustering increases from 0 to 5. The
reason for this is as follows: most of the interactions between cells
involve calculations in which at least one of the cells is sparsely
populated and only a few involve interactions between two densely
populated cells.  Although those sparsely populated cells finish their
interactions soon, they still remain active in the calculation since
all cells are processed in parallel. (The calculation finishes when
the densest cell has been processed.)  Clearly, one could avoid this
load-balancing problem by processing in parallel cells which have
about the same number of particles in the Top segments and hence all
cells processed in parallel finish their interactions at the same
time. As the algorithm is set-up to interact cells, this step is
rather easy to implement and the load-balanced algorithm is indeed
much more efficient, as Table 3 testifies.

\vbox{\centerline{\bf Table 3}
\vskip10pt
\centerline{\bf Load-balancing of PP part ($N_T=128k$)}
\tabskip=1.5em plus1em minus1em
\halign to\hsize{\hfil #\hfil &\hfil #\hfil&\hfil #\hfil & \hfil #\hfil &\hfil
#\hfil\ &\hfil #\hfil&\hfil #\hfil\cr
\noalign{\vskip10pt\hrule\vskip6pt\hrule\vskip6pt}
Load-balanced&$\sigma$&$N_{max}$&set-up
1&intra-cell&set-up 2&inter-cell\cr
\noalign{\vskip6pt\hrule\vskip6pt}
no&0.0&	4&	1.40&	0.07&		12.2&		10.7\cr
no&0.8&	10&	0.88&	0.29&		16.1&		30.8\cr
no&1.8&	30&	0.76&	0.69&		15.3&		65.3\cr
no&5.0&	204&	0.92&	4.30&		15.0&		353.0\cr
\cr
yes&0.0&	4&	1.10&	0.07&		11.8&		10.6\cr
yes&0.8&	10&	0.77&	0.19&		19.2&		19.0\cr
yes&1.8&	30&	0.77&	0.59&		29.1&		26.9\cr
yes&5.0&	204&	0.83&	4.10&		43.5&		38.0\cr
\noalign{\vskip6pt\hrule}
}
\smallskip}

\secttitle{4. Summary and conclusions}

We have presented a parallel P3M algorithm which corrects exactly for
inaccurate grid-induced PM forces between neighbouring particles.  The
PM step uses FFT's to compute the gravitational force between distant
particles and convolution of the density with the central part of the
Green's function to subtract grid-induced forces between neighbouring
particles in a time $\propto N_T$. Although this corrections step
requires a large number of operations to be executed on the grid, the
algorithm is still economically viable since these parallel statements
are executed extremely fast on the CM2 (running at $\approx
500-600$~MFlops). The PP step uses parallel processing of cells and
duplication of particles to efficiently compute neighbour forces in
situations where the particle distribution is clustered. The PP step
uses the same boundary conditions (i.e. isolated or periodic) as the
PM step does.  A full force calculation in a triply periodic system
for $128k$ particles on a $128^3$ grid with a rather high degree of
sub-clustering takes about 196 seconds (14 seconds for the PM
calculation, 95 for neighbour force subtraction and 86 for PP
calculation) on a $8k$ CM2.  Scaling the timings from the $8k$ nodes
CM2 to a $64k$ nodes CM200, this becomes an estimated 9.8 seconds.

\noindent {\it Acknowledgments: } I would like to thank Massimo
Stiavelli for comments on an earlier draft and Mark Rathsack for a
careful reading of this version of the paper and for his many
suggestions on the implementation of the P3M program.

\secttitle{Appendix. CM2 optimization considerations}

The Connection Machine 2 (CM2) is a massively parallel SIMD computer
consisting of thousands of simple processors, each having its own
private memory and interconnected with a hypercube topology.  The user
interacts with this hypercube through a front end computer which is a
conventional workstation. The computer at the Scuola Normale has 8192
processors and a total of 256MBytes of internal memory (i.e.  8192
single precision real numbers per processor).

The top speed of this configuration is $\approx $1Gflop. High
execution speeds (typically about half of the quoted maximum speed for
applications written in high level language) are obtained when
operations are done on operands which are in the local memory of each
processor (i.e., when no communication between processors is
necessary). In addition, the length of the vectors on which the
operations are done should be at least 3-4 times the number of
processors to amortize start-up costs. Figure 1 in Theuns and Rathsack
([10]) illustrates some typical execution speeds measured on this CM2.

CM Fortran contains some extensions that allow operations typical for
parallel computers.  One of these extensions, which we used
extensively in this implementation, is the CSHIFT function which
shifts all elements in an array over a specified number of positions
in a given direction (e.g. $-x$, $+z$).  This routine uses the
underlying hypercube topology and is very fast.  In addition, it
allows for an easy implementation of periodic boundary conditions in
both the PM and the PP part. Another routine which is essential for
the PP part is a functionality called scanning. The scan routine
allows the execution of an operation (e.g. copying, summing, etc.)
along a 1D array.  The scope of the operation is guided by logical
vectors which define segments in the array. The scan functionality
applies an operator to all values in a segment, and all segments are
scanned in parallel.  This routine allows for the copying of an
element in a cell to all other elements in another (or the same) cell
and for the summing of the contributions of those particles to the
copied one.  Finally, there is a very fast vendor supplied complex to
complex FFT on the CM2.

CM Fortran permits arrays to be allocated dynamically with a size
determined at run time.  This allows one to optimize the size of the
vectors on which the parallel operations are executed and in addition
save on memory since arrays which are not used in a given subroutine
can be deallocated. (For example, the grid arrays in the PM part are
not used in the PP part and hence can be deallocated.)  Consider for
example the PP step: initially we interact cells which have few
particles in the Top cell but potentially many in the Bottom cell: as
a result, vectors $B$ and $T$ are very large and their size in
practise is limited by the total memory available (512$k$ in our
case). Using such large vectors substantially reduces the
communication overhead costs involved in loading them. As the PP
calculation progresses, however, fewer and fewer cells are interacted,
yet each cell requires more and more interactions to complete. Hence,
efficiency is increased considerably by doing these interactions on
much smaller matrices.  Memory is freed at the end of a calculation
when the dynamically allocated vectors are deallocated.

\secttitle {References}

\noindent
[1] Aarseth, S.J.,in: Multiple time scales, Eds. J.U.Brackhill,
and\next\rs B.I. Cohen (Academic Press, New York, 1985) p. 377\next
[2] Hockney, R.W. and Eastwood, J.W. 1981, in: Computer Simulations
using Particles\next\rs (McGraw Hill, New York, 1981)\next
[3] James, R.A., Comput. Phys. 25 (1977) 71\next
[4] Efstathiou, G., Davis, M., Frenk, C.S., White, S.D.M., ApJS 57 (1985)
241\next
[5] Villumsen, J.V., ApJ 71 (1989) 407\next
[6] Appel, A.W., Thesis, Princeton University (1981) (unpublished)\next
[7] Barnes, J. and Hut, P., Nature 324 (1986) 446\next
[8] Hernquist, L., J. Comput. Phys 87 (1990) 127\next
[9] Makino, J. and Hut, P., Comp. Phys. Rep. 9 (1988) 199\next
[10] Theuns, T. and Rathsack, M.E., Comp. Phys. Comm. 76 (1993) 141\next

\secttitle {Figure Captions}
\noindent
Fig.1 :\next PM forces and potential on massless test particles
randomly distributed in a $16^3$ grid due to eight massive particles
randomly distributed in cell (8,8,8) as a function of the distance
$|{\bf r}|$ of the particle to vertex (8,8,8). Shown are the
deviations (in \%) of : (a) radial, (b) tangential PM force and (c) PM
potential with respect to their Newtonian counterparts in units of
total Newtonian force and Newtonian potential respectively.  Particles
in cells not neighbouring cell (8,8,8) are denoted by crosses, whereas
particles in cell (8,8,8) or one of it's neighbouring cells are
denoted by dots. For the latter particles, relative deviations have
been divided by four. The forces and potential on these particles will
be exactly equal to their Newtonian values after correction step and PP
part.\next

\bigskip
\noindent
Fig. 2a:\next Illustration of the intra-cell calculation. Particles in
the same cell are indicated with the same letter
($a$,$b$,$c$,$\cdots$) and particles in a given cell are numbered
($a_1$, $a_2$, $\cdots$).  In the top panel, the Top array $T(1)$ is
identical to $B$ but shifted to the left by one.  Force calculation
now occurs between different particles that are in the same cell:
these pairs are underlined.  Next, $T$ is shifted once more and again
forces are calculated.  The process finishes in step 4, when there are
no more particles in $T$ that interact with particles in $B$.\next

\bigskip
\noindent
Fig.  2b:\next Illustration of the inter-cell calculation. Particles
in the Bottom array are designated with lower case letters and
particles in the Top array that belong to a neighbouring cell with the
corresponding uppercase letter. Segments are delimited by vertical
lines and empty array positions are denoted by a dot. The process
starts by copying, for each cell in the $T$ array, its first element
($A_1$, $B_1$, $\cdots$) into array $I$ and next copying this particle
into array $C$ over the whole segment delimited by $B$. Forces are now
calculated between all particles in $B$ which have their corresponding
uppercase particle above them in $C$. Next, $T$ is shifted to the left
and the process is repeated until all particles in $T$ have been
processed.
\end